\documentclass[prl,twocolumn,amsmath,amssymb,superscriptaddress,citeautoscript,longbibliography]{revtex4-2}

\usepackage{graphicx,color}
\usepackage{braket}
\usepackage[colorlinks,urlcolor=blue,citecolor=blue]{hyperref}

\newcommand{\Rcal}{\mathcal{R}}
\newcommand{\Qcal}{\mathcal{Q}}

\begin{document}
\title{Quantum coherent states of interacting Bose-Fermi mixtures in one dimension}

\author{J. Clayton \surname{Peacock}}
\affiliation{Department of Physics, New York University, New York, NY-10003, USA}
\affiliation{Department of Physics, University of Cincinnati, Cincinnati, OH-45221, USA}
\author{Aleksandar \surname{Ljepoja}}
\affiliation{Department of Physics, University of Cincinnati, Cincinnati, OH-45221, USA}\author{C.~J.~\surname{Bolech}}
\affiliation{Department of Physics, University of Cincinnati, Cincinnati, OH-45221, USA}
\affiliation{ITAMP, Harvard-Smithsonian Center for Astrophysics, Cambridge, MA-02138, USA}

\begin{abstract}
We study two-component atomic gas mixtures in one dimension involving both bosons and fermions.
When the inter-species interaction is attractive, we report a rich variety of coherent ground-state phases that vary with the intrinsic and relative strength of the interactions.
We avoid any artifacts of lattice discretization by developing a novel implementation of a continuous matrix-product-state \textit{Ansatz} for mixtures and priorly demonstrate the validity of our approach on the integrable point that exists for mixtures with equal masses and interactions (Lai-Yang model), where we find that the \textit{Ansatz} correctly and systematically converges towards the exact results.
\end{abstract}

\maketitle

Since the experimental achievement of quantum degeneracy in weakly interacting atomic gases, first for bosons \cite{Anderson1995,*Bradley1995,*Davis1995} and soon after for fermions and boson-fermion mixtures \cite{DeMarco1999,Truscott2001,*Schreck2001}, there has been a tremendous interest in the study of such systems.
They provide an ideally clean and controllable experimental platform for fundamental studies of unitary quantum gases and for their use in applications ranging from quantum simulation to metrology.
For the case of mixtures, the new experimental platform reignited a long-standing theoretical interest connecting back to $^4$He-$^3$He solutions \cite{Ebner1971,*Edwards1992} and related to superfluidity and phase separation; but now combined with new practical considerations, since mixtures turned out to be also a useful stepping stone for the evaporative cooling of fermions with otherwise vanishing $s$-wave cross sections (a method known as sympathetic cooling \cite{Truscott2001,*Schreck2001,*Aubin2005}).

While the initial theoretical studies of gas mixtures (with mixed statistics) considered three-dimensional (3D) gas clouds \cite{Viverit2000}, soon the interest expanded to include one-dimensional confinement (via elongated traps or optical lattices) \cite{Das2003,Albus2003,Cazalilla2003}.
Those early studies unveiled a rich Luttinger-liquid phenomenology with instabilities towards collapse, demixing and pairing depending on the sign and strength of the interactions \cite{Cazalilla2003}.
Subsequent studies paralleled the even greater theoretical activity attracted by two-component Fermi mixtures \cite{Cazalilla2005,*Oelkers2006,*Orso2007,*Hu2007,*Guan2007,*Feiguin2007,*Parish2007,*Liu2008,*Casula2008,*Zhao2008,*Feiguin2009,*He2009,*Kakashvili2009,*Wang2009,*Baarsma2010,*Orso2010,*Baksmaty2011,*Baksmaty2011a,*Sun2011,*Baksmaty2012,*Dalmonte2012,*Bolech2012,*Gubbels2013,*Wang2013,*Guan2013,*Trenkwalder2014,*Roscher2014,*Liu2016,*Mei2016,*Sundar2020,*He2020}, that accompanied the experiments as they also evolved from 3D-trapped clouds \cite{Regal2004,*Zwierlein2004,*Kinast2004,*Bartenstein2004,*Bourdel2004,*Partridge2005,*Zwierlein2005,*Zwierlein2006,*Partridge2006,*Partridge2006b,*Jag2014} to the realm of 1D confinement \cite{Liao2010,*Liao2011,*Pagano2014,*Olsen2015,*Revelle2016,*Yang2018}. A similar evolution in the experimental study of Bose-Fermi mixtures has the potential to be equally rich and interesting.

The physical modeling of (one-dimensional) Bose-Fermi mixtures of atomic gases with short-range (contact) interactions is well approximated by the generalized Lai-Yang Hamiltonian \cite{Lai1971},
\begin{eqnarray}
H_{\textsc{ly}} & = & H_{\textsc{kin}}+H_{\textsc{int}}\,, \\
H_{\textsc{kin}} & = &
\int_{0}^{L}dx\sum_{\alpha=\textrm{b,f}}\frac{\hbar^2}{2m_{\alpha}}
\partial_{x}\hat{\psi}_{\alpha}^{\dagger}(x)
\partial_{x}\hat{\psi}_{\alpha}(x)\,, \\
H_{\textsc{int}} & = &
\int_{0}^{L}dx\sum_{\alpha,\beta=\textrm{b,f}}\frac{g_{\alpha\beta}}{2}
\hat{\psi}_{\alpha}^{\dagger}(x)\hat{\psi}_{\beta}^{\dagger}(x)
\hat{\psi}_{\beta}(x)\hat{\psi}_{\alpha}(x)\,,
\label{eq:LY}
\end{eqnarray}
where the fields $\hat{\psi}_{\alpha}^{\dagger}(x)$ create bosons or fermions with mass $m_{\alpha}$ (for the \textit{fermionicity} $\alpha\!\in\!\{\textrm{b\,=\,0}, \textrm{f\,=1}\}$, respectively) and obey the appropriate (anti)-commutation relations, $[\hat{\psi}_{\alpha}(x),\hat{\psi}^{\dagger}_{\beta}(y)]_{(\alpha\beta)}
=\delta_{\alpha,\beta}\,\delta(x-y)$; where $[A,B]_\alpha=AB-(-1)^\alpha BA$.
They interact with a contact-potential strength $g_{\alpha\beta}=2\hbar\omega_{\perp} a_{\alpha\beta}$, where $a_{\alpha\beta}$ is the corresponding scattering length between the two species and $\omega_{\perp}$ is the common transverse trapping frequency \cite{Olshanii1998}. 
Since fermions avoid each other, for contact interactions (with a range smaller than the exchange-correlation hole) the strength becomes arbitrary and we can set $g_\textrm{ff}=0$.

This model has been studied (away from integrable points) using mean-field and field-theoretical approaches (such as bosonization, in the hard-core-boson limit) \cite{Das2003,Cazalilla2003,Gautam2019}. However, most of the further studies of one-dimensional Bose-Fermi mixtures \cite{Albus2003,Lewenstein2004,*Rizzi2008,*Marchetti2009,*Singh2020,*Avella2020,*Guerrero2021} relied on lattice discretizations and mappings to the Bose-Fermi Hubbard model; plus the use of wave-function approximations such as the BCS mean-field, Gutzwiller and Jastrow \textit{Ans\"atze} \cite{Gutzwiller1963,*Krauth1992,*Sun2014,*Barfknecht2015,*Zhu2019}, or numerical methods such as the matrix-product-based density matrix renormalization group (DMRG) \cite{White1992,*White1993,*Kancharla2001}. 

During the last decade, a generalized coherent-state version of the matrix-product variational \textit{Ansatz} for quantum fields in a one-dimensional continuum was successfully put forward \cite{Verstraete2010,Maruyama2010,Osborne2010,Haegeman2010,Haegeman2013a,Draxler2013, Hubener2013,Quijandria2014, Stojevic2015,Chung2015,Quijandria2015,Mei2017,Draxler2017,Ganahl2017,Chung2017,Ganahl2018,Tilloy2019,Balanzo2020,Tang2020,Karanikolaou2021}.
This collective work explored the physics of a large number of systems of single or multi-component gases with a given type of statistics. 
We present here an extension of those developments that demonstrates an efficient implementation of the continuum matrix-product-state (cMPS) \textit{Ansatz} for a two-component gas with mixed statistics. 
We apply it to the study of the Lai-Yang model, first validating the method against the Bethe \textit{Ansatz} at the (non-trivial) integrable point and then studying the nature of pairing tendencies in certain regions of the non-integrable regime.

A cMPS for a binary (or multicomponent) gas mixture, for a system
of length $L$ and periodic boundary conditions, has the general form \cite{Haegeman2010}
\footnote{For integrable models solvable with the (non-nested) quantum inverse scattering method, the exact wave functions take the same form but with an additional projector inside the trace that fixes particle number and restores the $U(1)$ invariance \cite{Maruyama2010,Mei2017}. 
The exact eigenstates can thus be regarded as having a hidden concomitant coherent nature uncovered by cMPS.}
\begin{eqnarray}
\left|\chi\right\rangle & = &
\mathrm{Tr_{aux}}[\mathcal{P}e^{\int_{0}^{L}dx\left[\Qcal(x)\otimes \hat{I}+\sum_{\alpha}\Rcal_{\alpha}(x)\otimes\hat{\psi}_{\alpha}^{\dagger}(x)\right]}]\left|0\right\rangle,\label{eq:cmps}
\end{eqnarray}
where the complex-valued local matrices $\Qcal(x)$ and $\Rcal{_\alpha}(x)$ act on an auxiliary space of dimension $D$, called the \emph{bond dimension}.
In addition, $\alpha$ is the index for the atom type (b or f), $\hat{I}$ is the identity operator on the Fock space, $\mathrm{Tr_{aux}}$ is a trace over the auxiliary space, $\mathcal{P}e$ is a path-ordered exponential, and $\left|0\right\rangle$ is the bare vacuum state annihilated by all the field operators.
Henceforth, we will consider flat-bottom trapping potentials and adopt a phase-modulated uniform \textit{Ansatz} \cite{Chung2015} given by $\Qcal(x)=Q$ and $\Rcal_{\alpha}(x)=e^{iq_{\alpha}x}R_{\alpha}$, where $Q$ and $R_{\alpha}$ are three position-independent matrices, and $q_{\alpha}$ are two additional real-valued variational parameters (continuum in the large-$L$ limit).

For regularity, the $R_{\alpha}$ have to obey the same local algebra as the matching field operators, $[R_{\alpha},R_{\beta}]_{(\alpha\beta)}=0$. 
In particular, nilpotency in the fermion sector, $R^2_\textrm{f}=0$, has to be implemented exactly for improved numerical stability \cite{Chung2015}. 
Thus, ignoring zero eigenvalues, $R_\textrm{f}$ has to have an \textit{unrescaled} Jordan form 
\footnote{This is a particular case of what is know as a \textit{quasi} Jordan (canonical) form \cite{Golub1976}. 
Notice that our usage of the Jordan form is purely algebraic, to enforce second-degree nilpotency, and we are thus not concerned with the notorious numerical issues presented by \textit{defective} matrices.}
composed entirely of $2\!\times\!2$ blocks with zero Jordan eigenvalue (i.e.,~proportional to the spin-rising Pauli matrix, $\sigma^+$); notice this restricts $D$ to take only even values. 
The inclusion of scaling parameters in the Jordan blocks provides better convergence for mixtures with (very) different densities of each species.
Explicitly, one can write $R_\textrm{f}=\sigma^+\otimes\Gamma$ and enforce commutation with its boson counterpart by taking $R_\textrm{b}=\sigma^\uparrow\otimes A_\textrm{b}+\sigma^+\otimes B_\textrm{b}+\sigma^\downarrow\otimes D_\textrm{b}$ where $A_\textrm{b}, B_\textrm{b}, D_\textrm{b}, \Gamma\in\mathbb{C}^{D/2\times D/2}$ and $\sigma^\sigma=\left|\sigma\right>\left<\sigma\right|$ are Pauli projectors.
The enforcement requires $\Gamma D_\textrm{b}=A_\textrm{b}\Gamma$, which we do by solving for $D_\textrm{b}$ and keeping the other matrices arbitrary for variational optimization
\footnote{If $\Gamma$ is singular, we can define $D_\textrm{b}=\Gamma^+A_\textrm{b}\Gamma$ using the Moore-Penrose pseudoinverse. 
This reintroduces the possibility of zero eigenvalues in $R_\textrm{f}$, but we found that it was not necessary in practice.}
\footnote{We found that the cMPS \textit{Ansatz} for two-fermion mixtures can be re-implemented more efficiently based on this Bose-Fermi one; the details will be given elsewhere.}.

The cMPS norm is given by $\left<\chi|\chi\right>=\mathrm{Tr}\left[e^{TL}\right]$ where $T=T_+$ and $T_\pm=\bar{Q}\otimes I+I\otimes Q+\bar{R}_{\textrm{b}}\otimes R_{\textrm{b}}\pm\bar{R}_{\textrm{f}}\otimes R_{\textrm{f}}$ (the bars denote complex conjugation of matrix entries).
Noticing that the states are invariant under arbitrary similarity transformations of the matrices (that leave the trace invariant), one identifies the possibility of making the \textsl{gauge-fixing} choice $Q^\dagger+Q+R_{\textrm{b}}^\dagger R_{\textrm{b}}+R_{\textrm{f}}^\dagger R_{\textrm{f}}=0$. 
That guarantees a cMPS has unit norm in the thermodynamic limit \cite{Verstraete2010,Haegeman2013a}.
We can then have a \textsl{right-identity-normalization} of $T$ by taking $Q=A-\frac{1}{2}R_{\textrm{b}}^{\dagger}R_{\textrm{b}}-\frac{1}{2}R_{\textrm{f}}^{\dagger}R_{\textrm{f}}$, where $A$ is an arbitrary anti-Hermitian matrix
\footnote{The null right eigenvalue of $T$ is given by the identity matrix (with the double index of the direct-product basis interpreted as rows and columns, respectively).}. 
We find that a numerically accurate normalization of $T$ is important for the convergence properties of the algorithm.

The $\alpha$-atom number density can be readily computed as $n_\alpha=\bigl\langle\hat{\psi}_{\alpha}^{\dagger}(x)\hat{\psi}_{\alpha}(x)\bigr\rangle=\mathrm{Tr}[e^{TL}(\bar{R}_{\alpha}\otimes R_{\alpha})]$ and similarly for the corresponding kinetic-energy density by using $\bigl\langle\partial_{x}\hat{\psi}_{\alpha}^{\dagger}(x)\partial_{x}\hat{\psi}_{\alpha}(x)\bigr\rangle
=\mathrm{Tr}[e^{TL}\{\mathrm{c.c.}\otimes(iq_{\alpha}R_{\alpha}+\left[Q,R_{\alpha}\right])\}]$.
Interaction terms and correlators admit similar (longer) expressions in which the $R_{\alpha}$ matrices replace the field operators \cite{Verstraete2010,Haegeman2013a,Chung2015}.

\begin{figure}[t]
\includegraphics[trim=5 5 45 40,clip,width=8.6cm]{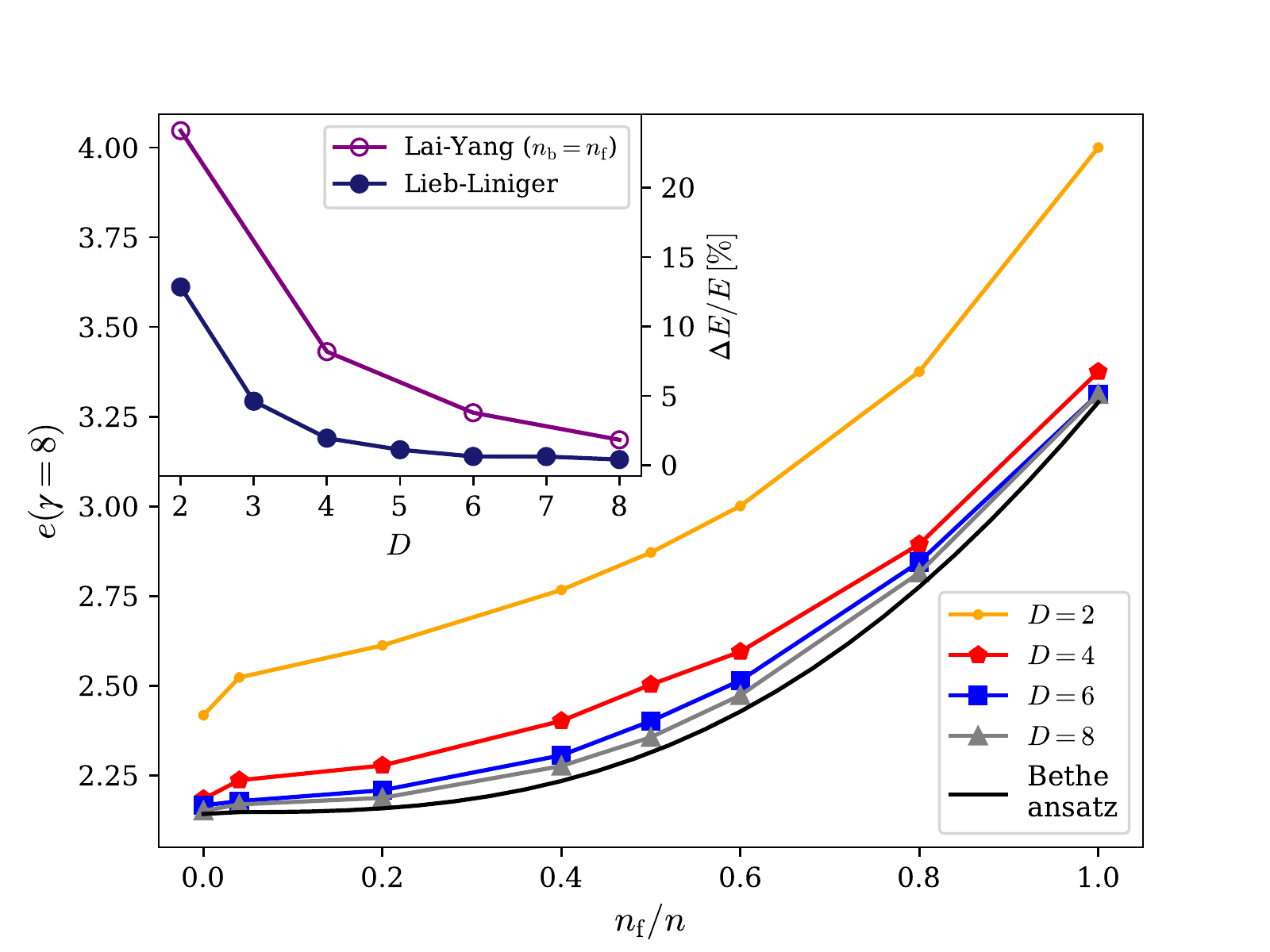}
\caption{\label{fig:LaiYang}
Variational cMPS ground-state energy, $e(\gamma)=2mE/(n^3\hbar^2L)$, for the Lai-Yang model with different bond dimensions as a function of the fermion-density fraction and compared with the Bethe \textit{Ansatz} result.
Inset: cMPS convergence with bond dimension of the relative energy error, $\Delta E=E_\textrm{c\sc{mps}}-E$, as a percentage of the Bethe-ansatz value, for $n_\textrm{f}/n=0.5$ and $0$ (the latter is the Lieb-Liniger limit that we computed with a bosons-only cMPS that has more variational parameters for a given $D$ and yields lower energies).}
\end{figure}

In order to demonstrate the accuracy of the cMPS \textit{Ansatz}, we focus first on the Lai-Yang integrable point \cite{Yang1967,Sutherland1968,Lai1971,Lai1974}, which corresponds to equal masses ($m_\textrm{b}=m_\textrm{f}=m$) and interactions ($g_\textrm{bb}=g_\textrm{bf}=g_\textrm{fb}=g>0$
\footnote{The attractive case ($g<0$) is also integrable but the system is expected to be unstable to the formation of solitons, as in the Lieb-Liniger case \cite{Cuevas2013,*Yurovsky2017}}) between bosons and fermions. 
In experiments, the first condition can be approximately fulfilled by considering different isotopes of the same atom, and the second one by tuning the interactions via shape and Feshbach resonances (deviations break integrability, but can still be modeled within cMPS \cite{Chung2017}). 
There have been several studies of this special case using the Bethe \textit{Ansatz} and framed in the modern context of cold-atom gases \cite{Imambekov2006,*Imambekov2006a,Batchelor2005,*Guan2008,Frahm2005,Hu2016}. They found that, despite earlier suggestions, the demixing tendency is absent for the homogeneous system (but one can still have trap-induced Bose-Fermi separation; cf.~Ref.~\onlinecite{Imambekov2006b}). 
We will not explore those questions here; rather, we will focus on comparing variational ground-state energies with the exact results and assessing the convergence properties of cMPS; see Fig.~\ref{fig:LaiYang}.
We will work with units in which $\hbar=1$ and $m=1$ \footnote{Also $2m=1$ is another popular choice; cf.~Ref.~\onlinecite{LiebLiniger1963,*Lieb1963}}, will take $n=n_\textrm{b}+n_\textrm{f}=1/4$ per unit length (in some arbitrary units), and the interaction strength will be given by setting the dimensionless parameter $\gamma=g/n=8$ \footnote{Middle of the range considered in Fig.~1 of Ref.~\onlinecite{Imambekov2006,*Imambekov2006a}}.

We observed good systematic convergence of the cMPS energies as a function of bond dimension---as expected, since the \textit{Ansatz} captures progressively more and more multi-particle entanglement. 
For two-component Bose-Fermi mixtures, convergence in $D$ is slower than in the Lieb-Liniger case, but faster than for the Gaudin-Yang model (the two-fermion \textit{Ansatz} can be constructed as an outer nesting layer on the one given here and is further restricted to $D$ being a multiple of $4$; cf.~Ref.~\onlinecite{Chung2015}).

We minimized the energy of the system directly in the \mbox{$L\to\infty$} limit and tested several standard local-optimization algorithms. We found that a principal-axis minimization was usually the best strategy \cite{Brent1972}, although other ones, such as the Nelder-Mead simplex method, also converged well. 
The global optimization can be done using simulated annealing, but we found that repeated random starts and parametric variations provided a simpler strategy and gave comparable results for not-too-large bond dimensions. 
We targeted fixed particle densities using (augmented) Lagrange multipliers and penalties. 
The single-optimization running times on regular hardware ranged from a few seconds for \mbox{$D=2$} to several hours for \mbox{$D=8$}.

Having established the viability of the cMPS \textit{Ansatz} for the study of mixtures, we next move on to address the case of regimes with attractive interactions between bosons and fermions, $g_{\textrm{bf}}=g_{\textrm{fb}}=-g<0$.
All the while, the interactions between bosons shall remain repulsive, $g_{\textrm{bb}}/g=G>0$, in order to guarantee the thermodynamic stability of the system by preventing \textit{bosonic collapse}. The large-$G$ limit corresponds to the Tonks-Girardeau regime \cite{Tonks1936,*Girardeau1960} in which the effective exclusion statistics of the bosons gradually approximates that of fermions and the low-energy spectrum of the model converges to that of a Gaudin-Yang gas.
One would thus expect a tendency to (algebraic) superconducting order as in the two-fermion case; however, while some of this intuition is correct, the exchange statistics of the bosons remains bosonic (they tend to \textit{hard-core bosons}) and the emerging physical picture is considerably more subtle, as shall be established below.

\begin{figure}[t]
\includegraphics[trim=5 38 45 40,clip,width=8.6cm]{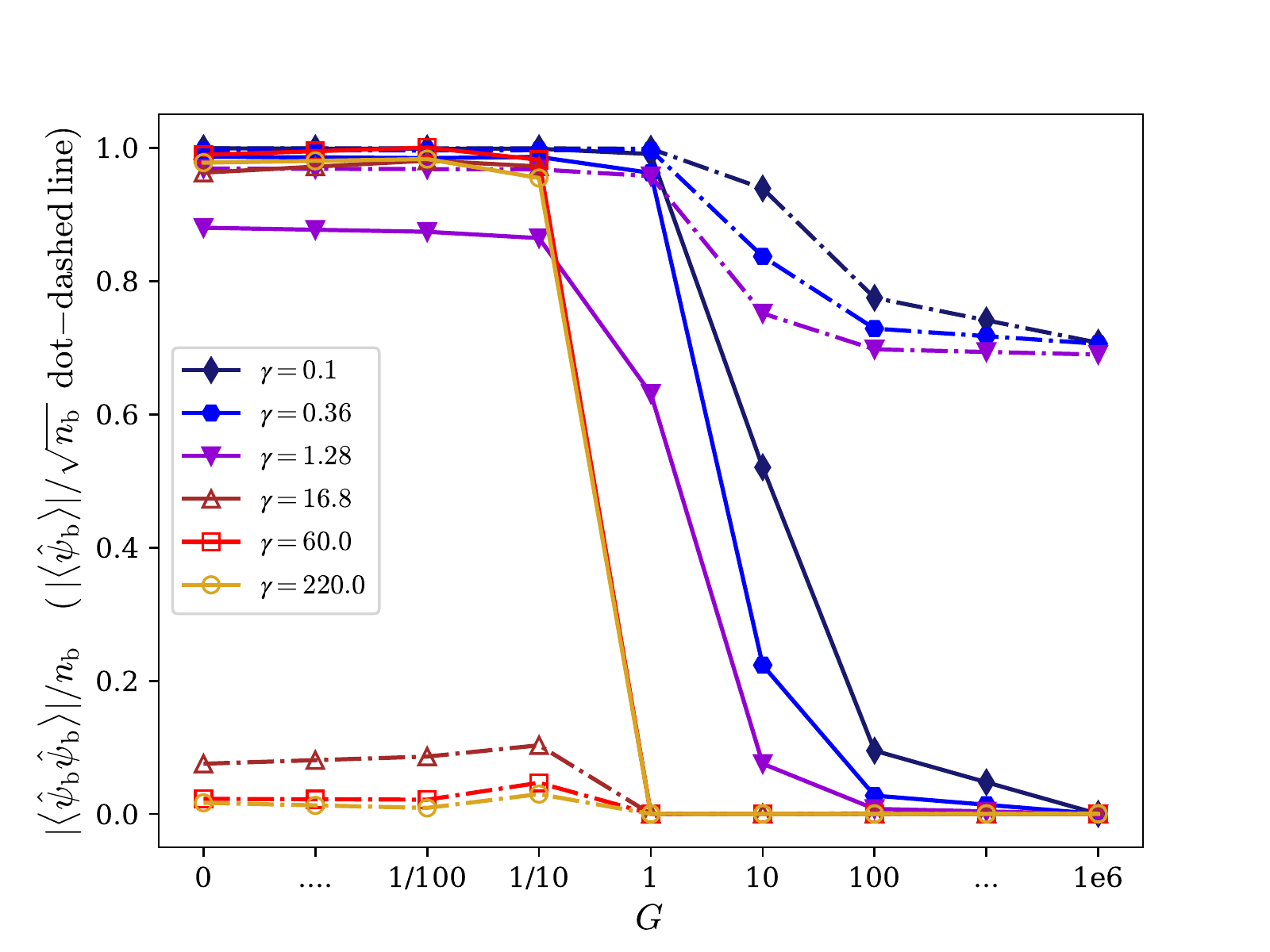}
\includegraphics[trim=5 38 45 40,clip,width=8.6cm]{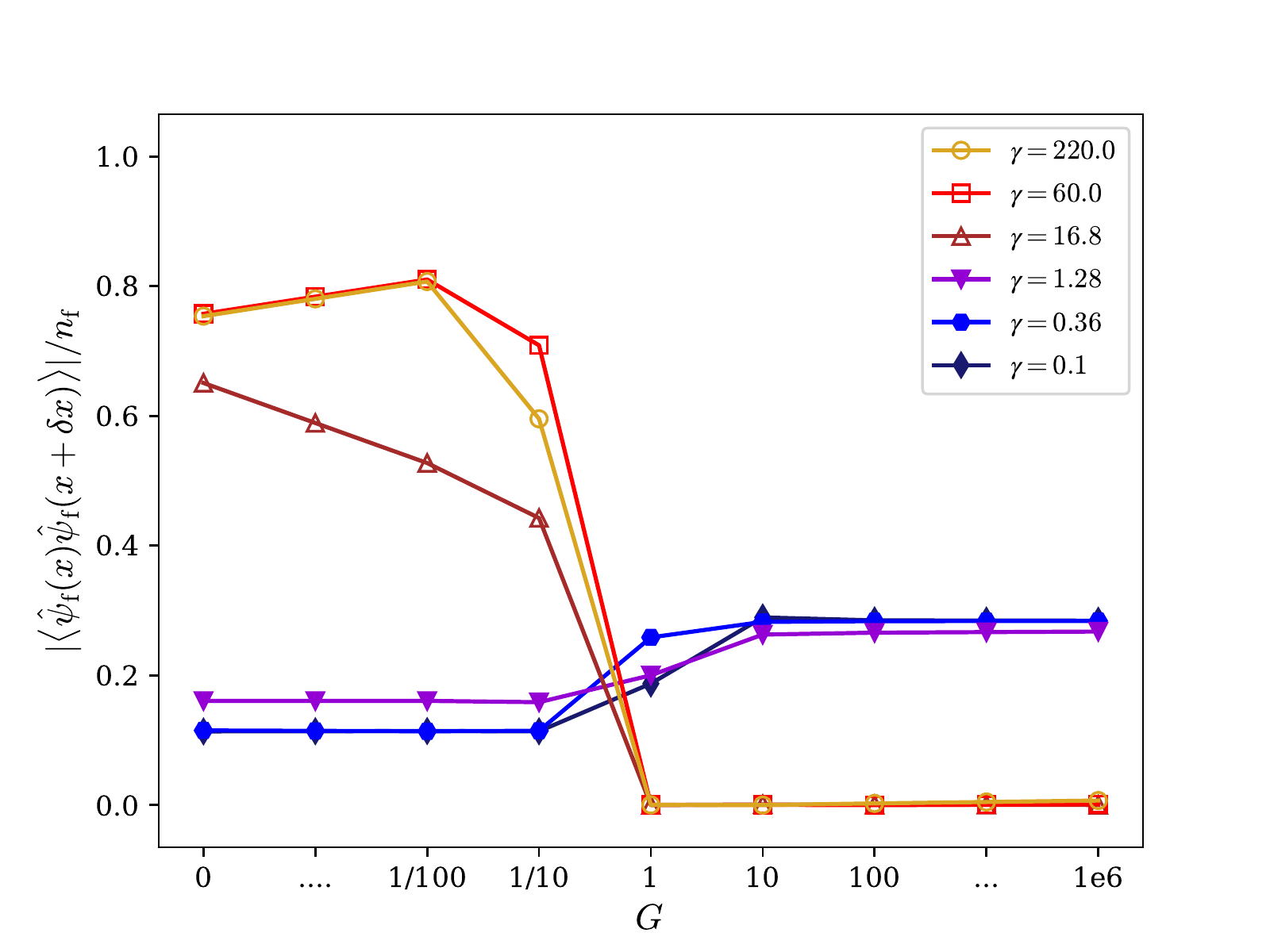}
\includegraphics[trim=5  5 45 40,clip,width=8.6cm]{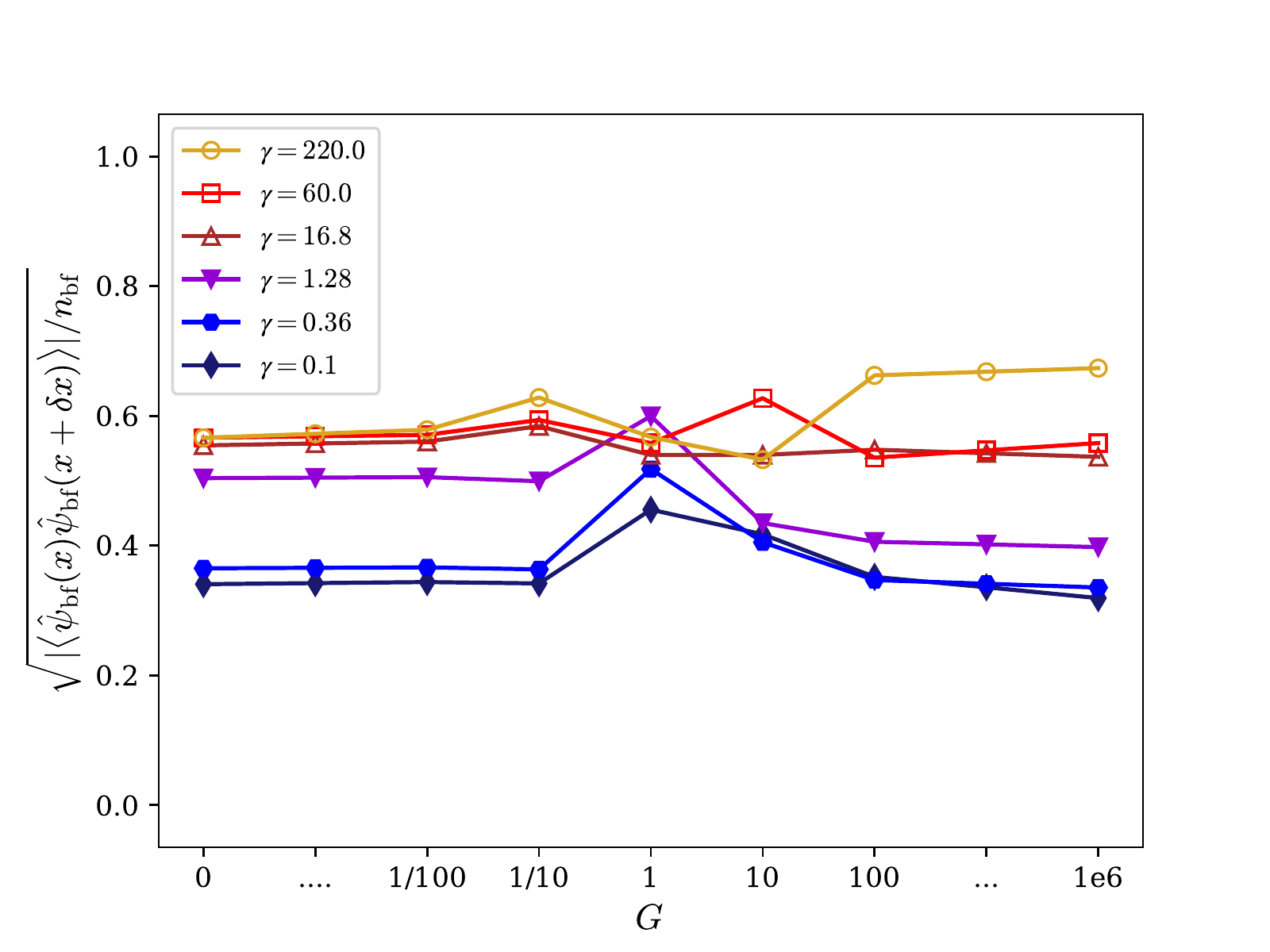}
\caption{\label{fig:OrderParameters}
Normalized coherence order parameters for a Lai-Yang gas with repulsive inter-boson and attractive boson-fermion interactions. 
The calculations were done using $D=4$ cMPS states and the typical error is expected to be in the 5-10\% range. Contrasting behaviors are clearly seen between the strongly 
($\gamma\gg1$) and weakly ($\gamma\lesssim1$) interacting regimes.}
\end{figure}

Our implementation of the cMPS \textit{Ansatz} constitutes a \textit{sui generis} matrix representation of generalized coherent states, and as such it is particularly well suited to capture the emergence of ground states with spontaneously broken $U(1)$ symmetries and off-diagonal (quasi) long-range order. 
For instance, bosonic field operators can acquire a non-zero vacuum expectation value (vev), $\langle\hat\psi_\textrm{b}(x)\rangle$, that signals the occurrence of (quasi) Bose-Einstein condensation (BEC), a state with macroscopic quantum phase coherence \cite{Bogolyubov1947}. 
We shall refer to the absolute values of such appropriately normalized vevs as \textit{coherence order parameters} (cf.~Ref.~\onlinecite{Glauber1963}). 
They take values in the unit interval and a comprehensive set of them is displayed in Fig.~\ref{fig:OrderParameters} for a wide range of values of $\gamma=g/n$ and $G$.

Due to their statistics, fermions cannot condense as single atoms, but they can pair up and condense as a molecular BEC (localized pairs), or a more BCS-like state (extended pairs), or any intermediate scenario (BEC-BCS crossover). 
Such pair-coherent states (also possible for bosons) are signaled by the vevs $\langle\hat\psi_\alpha(x)\hat\psi_\alpha(x+\delta x)\rangle$, where $x$ is arbitrary due to translation invariance and $\delta x$ is chosen to maximize the amplitude (and corresponds to the most likely atom-atom distance in the pair, which has to be nonzero for spinless fermions due to exclusion but turns out to be zero for bosons).
Na\"{\i}vely, one would want to consider also mixed-species pairs, but their combined Fermi-Dirac statistics precludes condensation (even in the Tonks-Girardeau regime). 
Rather, Bose-Fermi coherence manifests in composite order parameters (cf.~Refs.~\onlinecite{Mathey2004,*Mathey2007a,*Mathey2007b,Marchetti2009}).
After exploring different correlators, we are led to define the local bf-\textit{molecule} field $\hat\psi_\textrm{bf}(x)\equiv\hat\psi_\textrm{b}(x)\hat\psi_\textrm{f}(x)$ and consider its \textit{pair-of-pairs} (PoP) vev defined as above but with $\alpha\to\textrm{bf}$.

\begin{figure}[t]
\includegraphics[width=8.6cm]{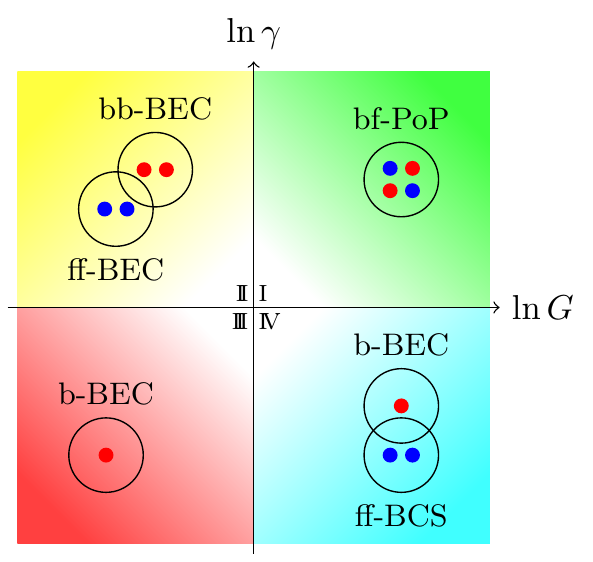}
\caption{\label{fig:phases}
Schematic representation of the different quasi-long-range-order ground-state phases present when the interactions among bosons are repulsive while bosons and fermions attract each other, as a function of interaction strengths.
The dominant orders are indicated in each case.}
\end{figure}

The combined consideration of all the order parameters defined above, yields a comprehensive picture of a ground-state phase diagram with four starkly different sectors and interpolating crossover regions between them.
This information is schematically summarized in Fig.~\ref{fig:phases} to guide the discussion.
Rather than comparing algebraic exponents as in bosonization-based studies, cMPS forces the breaking of $U(1)$ symmetries and allows the direct comparison of coexisting coherence-order-parameter amplitudes (normalized so that the dominant phases can be identified by their larger relative magnitudes in Fig.~\ref{fig:OrderParameters}).
The simplest case is the weakly interacting limit deep into the third quadrant of Fig.~\ref{fig:phases}, in which $\langle\hat\psi_\textrm{b}\rangle$ is the dominant order (approximately saturating the bound) signaling boson condensation.
Notice in the top panel of Fig.~\ref{fig:OrderParameters} that $\langle\hat\psi_\textrm{b}\hat\psi_\textrm{b}\rangle$ is also comparably large, as expected for true higher-order coherence \cite{Glauber1963}, while the other order parameters involving fermions are highly suppressed in comparison.
Increasing $G$ and moving into the fourth quadrant, the amplitude of the $\langle\hat\psi_\textrm{b}\rangle$ order parameter is partially suppressed and the coherence reduces to first-order only.
On the other hand, the $\langle\hat\psi_\textrm{f}\hat\psi_\textrm{f}\rangle$ order is enhanced in the process, with extended BCS-like pairing, and we find a scenario of combined coherence (compatible with the polaronic picture of earlier studies \cite{Mathey2004,*Mathey2007a,*Mathey2007b}).

If, instead, we keep $G$ small and increase $\gamma$, going into the second quadrant, the single-boson coherence is suppressed to very low values while the two-boson one remains close to maximal, and in addition the two-fermion coherence is greatly enhanced and more BEC-like than in the fourth quadrant (see the middle panel of Fig.~\ref{fig:OrderParameters}).
This is interpreted as Bose-Fermi \textit{mutualism}, with the bosons providing the \textit{glue} for the fermionic pairing while, simultaneously, the fermions do the same for the bosonic pairing.
The two order parameters are not mixed (they involve different degrees of freedom), but neither condensate would be able to exist without the other.
Finally, when $G$ and $\gamma$ are both large, we go into the first quadrant of Fig.~\ref{fig:phases}.
We find that all of the purely bosonic or purely fermionic orders are suppressed to zero (see the top and middle panels of Fig.~\ref{fig:OrderParameters}), while only the mixed $\langle\hat\psi_\textrm{bf}\hat\psi_\textrm{bf}\rangle$ four-particle coherence remains present in the system (bottom panel of the same figure).
The latter corresponds to tightly bound Bose-Fermi molecules that condense in loosely bound pairs. 
Moreover, this PoP order is also enhanced in the intermediate-coupling region at the center of the phase diagram, where it coexists with a strength similar to that of the dominant orders from the other quadrants in a large mixed-coherence crossover region.

We have focused on the case of balanced populations of equal-mass Bose and Fermi atoms, but those conditions might well not be the most easily achievable in experiments.
We also considered the case of a 20\% density imbalance (both ways) and found that the phase diagram remains the same asymptotically while the location of the crossover boundaries shifts with the density ratio.
One would expect similar results as a function of mass imbalance and cMPS would be an ideal tool for that study (based on our past experience with Gaudin-Yang systems \cite{Chung2017}); however, since the space of parameters is large, we leave a focused study for the future, to be guided by the parametric choices dictated by experimental considerations.
Similarly to the case for two-fermion gases, a convenient setup might be a quasi one-dimensional array of tubes created by a transverse optical lattice.
This would turn the ordering tendencies of the system into true long-range order by exploiting the 1D-3D crossover and can have a number of additional experimental benefits \cite{Marchetti2009,Revelle2016,Sun2013}.
On the other hand, the inclusion of a (weak) longitudinal optical lattice would break translation invariance and open up the possibility of density-wave orders that do not exist in the continuum (as found in studies based on the Bose-Fermi Hubbard model).
The cMPS-ansatz implementation can be extended to that case as well, as has been done already for Lieb-Liniger gases \cite{Ganahl2018}. 
This carries the advantage of treating translation invariance, or lack thereof, without the additional umklapp scattering introduced by the lattice discretization needed for the use of standard MPS methods (which, moreover, require the truncation of the local bosonic Hilbert spaces at each lattice site and introduce biases in the capture of BEC order). Finally, our implementation of the particle statistics turns out to be quite elegant and simpler than in the MPS/DMRG setting, (cf.~Ref.~\onlinecite{White1992,*White1993,*Kancharla2001} and the large body of subsequent work).

\begin{acknowledgments}
This work was funded by the NSF (Grant No.~PHY-1708049) and at times used computing resources from the Ohio Supercomputer Center (OSC) \cite{OhioSupercomputerCenter1987}. In its initial stages, the work benefited from an extended visit to ITAMP (Harvard-Smithsonian Center for Astrophysics). We also acknowledge the hospitality of the IIP-UFRN (Brazil).
\end{acknowledgments}

\bibliography{bfbiblio}

\end{document}